\documentclass[a4paper,12pt]{article}
\pdfoutput=1
\usepackage[a4paper, bindingoffset=0.2in, left=1in, right=1in, top=1in, bottom=1in, footskip=.25in]{geometry}

\usepackage{times}
\usepackage{amsmath,amssymb}
\usepackage{graphicx,afterpage}
\usepackage[final]{pdfpages}
\usepackage{authblk}
\usepackage[nottoc]{tocbibind}
\usepackage{setspace}
\usepackage{url}
\usepackage{hyperref}
\usepackage{xcolor}
\hypersetup{
  colorlinks   = true, %Colours links instead of ugly boxes
  urlcolor     = red, %Colour for external hyperlinks
  linkcolor    = blue, %Colour of internal links
  citecolor   = blue %Colour of citations
}
\usepackage[labelsep=period, labelfont=sc]{caption}
\usepackage{tabularx,ragged2e,booktabs}
\usepackage{sectsty}

\usepackage{caption}
\usepackage{lineno}

\usepackage{enumitem}
\title{Lost in Diversification}

\author[1]{\small Marco Bardoscia\footnote{Any views expressed are solely those of the author(s) and so cannot be taken to represent those of the Bank of England or to state Bank of England policy.}}
\author[2]{\small Daniele d'Arienzo} 
\author[3,4]{\small Matteo Marsili} 
\author[5]{\small Valerio Volpati}

\affil[1]{\footnotesize Bank of England, Threadneedle St, London EC2R 8AH, UK}
\affil[2]{\footnotesize Bocconi University, Department of Finance, Via Roentgen 1, 20136, Milan, Italy}
\affil[3]{\footnotesize The Abdus Salam International Center for Theoretical Physics, Strada Costiera 11, 34151 Trieste, Italy}
\affil[4]{Istituto Nazionale di Fisica Nucleare (INFN), Sezione di Trieste, Italy}
\affil[5]{Capital Fund Management, 23, Rue de l'Universit\'e 75007 Paris, France}

\date{\today}

\begin{document}
\maketitle

\begin{abstract}
As financial instruments grow in complexity more and more information is neglected by risk optimization practices. 
This brings down a curtain of opacity on the origination of risk, that has been one of the main culprits in the 2007-2008 global financial crisis. We discuss how the loss of transparency may be quantified in bits, using information theoretic concepts. We find that {\em i)} financial transformations imply large information losses, {\em ii)} portfolios are more information sensitive than individual stocks only if fundamental analysis is sufficiently informative on the co-movement of assets, that {\em iii)} securitisation, in the relevant range of parameters, yields assets that are less information sensitive than the original stocks and that {\em iv)} when diversification (or securitisation) is at its best (i.e. when assets are uncorrelated) information losses are maximal. We also address the issue of whether pricing schemes can be introduced to deal with information losses. This is relevant for the transmission of incentives to gather information on the risk origination side. Within a simple mean variance scheme, we find that market incentives are not generally sufficient to make information harvesting sustainable. 
\end{abstract}

%\tableofcontents

\section{Introduction}

Financial innovations have been seen as a formidable tool to increase the efficiency of the market, by controlling the risk of financial assets thus easing resource allocation between investors and the real economy. Yet, several authors  \cite{haldane2011systemic,battiston2012liaisons,caccioli2009eroding} have suggested that the increasing complexity of financial products may trigger the emergence of instabilities and systemic risks. The most commonly believed determinant of the 2007-2008 global financial crisis is the rise of structured financial products  \cite{coval2009economics}. The formidable complexity of these type of products effectively brought down a curtain of opacity between the risk originators and the buyers of the financial products, that hid the true risks of the underlying assets (e.g.\ mortgages, loans, credits) \cite{Ghent2014complexity,rajan2015failure}.
While the dangers of these instruments had been highlighted well before the crisis \cite{rajan2005}, most of the response to the crisis didn't address the core issue of transparency loss implicit in financial transformations, but rather focused on {\em ring fencing} the financial system with various measures \cite{Haldane2018}. An exception to that, is the proposal \cite{Ali2012towards} to build an efficient and standardized system, or a common language, through which information on the origin of risks should be easily available to all market participants. Such a \emph{financial barcode}, which might be attached to any financial product, should contain all the information that is relevant in order to make realistic estimates about return and risk of the product, from the risk profiles of the building blocks to the market fundamentals. Yet, it is not clear how such barcodes should be constructed, which information they should contain and whether they should be statically or dynamically updated, when new information is available. In particular, an interesting open question is whether demand for such barcodes may "naturally" arise and how barcodes should be priced, since without a barcode price the sellers would have no incentive for sharing the information.

Apparently, within the prevailing {\em market efficiency hypothesis} paradigm, according to which prices of any stock exchanged in the market reflects faithfully any relevant information \cite{fama1965behavior,bouchaud2017have}, these barcodes would be worthless. Indeed, for example, the price of Asset Backed Securities (ABS) were computed only on the basis of default probabilities of the underlying assets (e.g.\ mortgages). Even though, in principle, all the documentation about the underlying assets were available to buyers, the prices of ABS didn't depend at all on it, with the consequence that incentives for due diligence in collecting information on the underlying by issuers were lacking \cite{rajan2005,rajan2015failure}. Yet, market information efficiency resides on the balance between traders seeking information (fundamental analysis) and traders exploiting it (technical analysis), as shown by a wealth of results in agent based modelling of financial markets (see e.g.\ \cite{LuxMarches}). The former profit by the fact that the information they gather grants them an excess return. Here the profits of collecting information accrue to the individual trader, while in the case of a structured financial product these are passed over to the buyer. Accordingly they should be reflected in the price. This simple logic is the basis of the present paper.

As a first step, we quantify the transparency loss by the amount of bits of information \emph{lost in diversification}. Secondly, we address the issue of deriving the optimal barcode, the one that contains the maximal information on the return of the financial instrument. Then we compute the price associated to the barcode as the value of the information within a simple mean variance framework. The information loss and the barcode price are then quantified within a model system based on Gaussian variables (see \cite{tesiValerio} for an extension to binary variables). Within this framework we find that financial transformation implies large information losses and that market incentives are not generally sufficient to make information harvesting sustainable. 

The remaining sections are organized as follows: in Section 2 we discuss the general setting, information and financial transformation. Then we quantify information losses for a simple model of Gaussian log-returns and address the issue of information pricing. We conclude with some general remarks.

\section{The general framework}

Let's suppose we have a pool of $n$ assets, e.g.\ stocks, loans or mortgages, and let $\vec X = ( X_1, \ldots ,X_n)$ be the associated vector of (log-)returns. The values $\vec X $ are unknown to the inverstor, so we shall treat them as a vector of random variables, described by a probability distribution $p(\vec X)$. We consider a situation where some {side information} related to the stock $X_i$, e.g. the income of the borrower of the loan or information on the fundamentals of asset $i$, is possibly available. This information is captured by a random variable $Y_i$ that, inspired by Ref.\ \cite{Ali2012towards}, we shall call the {\em barcode} associated to asset $i$. Barcodes allow investors that bought the asset, to retrieve all information that is relevant to estimate the return of the asset, in the sense that, given the barcodes $\vec Y$, they can use the conditional distribution $p(\vec X | \vec Y)$ instead of $p(\vec X)$. 
%Both the values of $\vec X $ and the values of $\vec Y = ( Y_1, \ldots ,Y_n )$ are unknown to the inverstor, therefore they can be treated as vectors of random variables, described by a probability distribution $p(\vec X, \vec Y)$.
%The investor faces the decision problem of how to evaluate the uncertain future log-returns $\vec X$. He/she can perform this task either using a prior belief, encoded in the marginal distribution $p(\vec X) $, or buying the additional information $\vec Y$ from the portfolio manager and using instead the conditional distribution $p(\vec X | \vec Y)$. We are interested to understand in which cases the retrieval of the information is relevant for the risk assessment.} 
% THIS IS A QUESTION WE ADDRESS LATER, THAT OF THE PRICING OF INFORMATION
We shall use the mutual information \cite{Cover} 
\begin{equation}
\label{eq:mi}
I(U,V)=\mathbb{E}\left[\log\left[\frac{p(u|v)}{p(u)}\right]\right] ,
\end{equation}
to quantify the amount of information that the knowledge of a variable $V$ provides on the random variable $U$. In Eq.\ (\ref{eq:mi}) $p(u|v)$ is the conditional distribution of $U$ given that $V=v$ and $p(u)$ is the unconditional one\footnote{For continuous variables $p(u|v)$ and $p(u)$ are probability density functions, see e.g.\ \cite{Cover}.}.  
Hence, $I(X_i,Y_i)$ measures in bits the information that $Y_i$ provides on $X_i$ and $I(\vec X,\vec Y)$ measures the total amount of information that the barcodes  $\vec Y$ provide on the returns $\vec X$. 
%\textcolor{red}{It is often useful to quantify how important the amount of information contained in $I(U,V)$ is compared to the bits necessary to specify the random variable $U$. For this reason we introduce also the mutual information \emph{per bit}
%\begin{equation}
%I(U,V)/H(U)
%\end{equation}
%}

\subsection{Financial transformations}

We consider financial transformations
\begin{equation}
\vec X\to F(X),\qquad 
X=\sum_{i=1}^n X_i,
\end{equation}
that entail pooling the $n$ assets into a single portfolio $X$ and applying a transformation $F(X)$. This generates a new financial asset with log-return $F(X)$. The simplest such transformation is the  portfolio itself that delivers the average log-return 
\begin{equation}
F_{\bar{X}}(X)\equiv\overline{X}=X/n.
\end{equation}
Here, $\vec X$ can be the log-returns of individual stocks. In this case, $\vec Y$ would encode information on fundamentals (e.g.\ corporate structure of the firm, analysis of the sector they operate etc) for each stock. 

$\overline{X}$ corresponds to the most basic diversification techniques, which entails investing a fraction $1/n$ in each of the $n$ assets, instead of investing in a single asset $X_i$. The benefit of diversification is that it reduces risk. For example, for $n$ i.i.d.\ stocks, the variance $\mathbb{V}(\overline{X})=\mathbb{V}(X_i)/n$ is reduced by a factor of $n$, w.r.t.\ that of individual stocks. 

Another class of products we consider are \emph{Asset Backed Securities} (ABS), the typical products of structured finance \cite{coval2009economics}, whose return function is based on a prioritized structure of claims. In these products, the claims over the cash flow of the returns of the underlying assets $X_i$ are structured in such a way that the ABS yields a positive return when the total return is larger then a given threshold $k$. The return of these products is
\begin{equation}
F_k (X) = \theta \left( X - k \right) ,
\end{equation}   
where $\theta (x) = 1$ when $x \geq 0$ and $\theta (x) = 0$ otherwise.
%\footnote{\textcolor{blue}{The Gaussian average log-return can be thought in this specification as a large $n$ approximation of a pool of loans, where each one can either default or not.}}. THIS COMMENTS RELATES TO THE GAUSSIAN MODEL WHICH WE DIDN'T YET INTRODUCE
Different tranches correspond to different risk profiles that can be obtained with different values of $k$. The transformation $\vec X \to F_k(X)$ is an example of securitisation and the advantage of it is that it turns a set of risky assets $X_i$ into assets with a controlled risk profile. Sufficiently small values of $k$ yield assets that are very safe, i.e.\ for which $F_k(X)=1$ with high probability. As an example, mortgage backed securities (MBS) \cite{coval2009economics} are based on a portfolio $\vec X$ of mortgages granted to $n$ households, where $X_i=+1$ if household $i$ repays the mortgage and $X_i=0$ if $i$ defaults. In this case $Y_i$ may encode the occupational status of $i$, the characteristics of the neighbourhood of the house bought with the $i^{\rm th}$ mortgage. In this case, $i$'s default may occur for idiosyncratic reasons, or for systemic ones (e.g.\ crisis in the sector of the economy of the company where Mr.\ $i$ works or a natural disaster in that region) that may affect different households in the same way. 

Investors can transmit all the information $\vec Y$ about the individual assets to the buyers of the engineered asset $F$. Yet, some of this information may not be relevant to estimate the return $F(X)$, i.e. all the information relevant to estimate the return of $F$ may be compressed in a single variable $G_F$ that we call the barcode of $F$. Clearly, $G_F (\vec Y)$ has to be a function  of  $\vec Y$, and, ideally, the barcode $G_F$ should be the simplest\footnote{Here simplest, in information theoretic terms implies the one requiring less bits for its description. For discrete variables this corresponds to the variable $V$ with the smallest entropy $H[V]=-\sum_V p(V)\log p(V)$. For continuous variables it is necessary to resort to the relative entropy $D_{KL}(p||p_0)=\int dV p(V)\log [p(V)/p_0(V)]$, where $p_0(V)$ is a baseline distribution. In the cases we shall discuss in the following the notion of simplicity is rather intuitive, so we shall not discuss these details further.} among all possible variables $V(\vec Y)$ such that $I\left(F(X),V(\vec Y)\right)=I\left(F(X),\vec Y\right)$. 

A general result  can be obtained by invoking the \emph{data processing inequality} \cite{Cover}. This states that in any transformation $\vec X \to F(X)$ some information may or may not be lost, but for sure no information can be gained. In terms of the mutual information this reads
\begin{equation}
\label{dpi:barcodes}
I(F(X),G_F(\vec Y)) \le I (F (X) , \vec Y) \leq I(X,\vec Y) \leq I (\vec X , \vec Y ).
\end{equation}
The term on the right end of this chain of inequalities quantifies the total amount of information in bits that barcodes $\vec Y$ provide on the log-returns $\vec X$. In the typical case of weakly dependent assets, this is proportional to the number $n$ of assets. By contrast, the second term from the right is upper bounded by the entropy of the random variable $X$, which grows at most as $\log n$. Hence, generally, financial transformations imply information losses. 
The choice of the optimal barcode $G_F(\vec Y)$ can only mitigate further information losses, and at most saturate the leftmost inequality in Eq.\ (\ref{dpi:barcodes}). 

%\textcolor{blue}{A general result on the optimal barcoding can be stated for the type of products we introduced before, i.e the portforlio and the ABS. Both these type of products are function of the asset log-returns $\vec X_i$ through their aggregate log-return $X$. The probability distribution for the aggregate log-return $X$ have the following property: 
%\begin{equation}
%p ( X \,| \vec Y ) = p ( X \,| Y ),
%\end{equation}
%where $Y$ is the aggregate information $Y = \sum_i Y_i$. Such a property holds as a consequence of the permutation symmetry of the sum. 
%This implies that the sum  $Y$ is an optimal barcode for any financial transformation $F(X)$ which is a function of the average log-return of the assets, i.e.
%\begin{equation}
%\label{eq:ResultBarcoding}
%I (F (X) , Y) = I (F (X) , \vec Y) .
%\end{equation}
%} IDON'T BELIEVE THIS RESULT IS TRUE

In the next section we shall turn to the quantitative analysis of a representative case.

\section{Barcoding finance in a Gaussian world}

Let's assume that 
\begin{eqnarray}
\label{eq:model}
X_i & = & \mu+ \xi_i+a\xi_0+JY_i \\
Y_i & = & \eta_i+c\eta_0 
\end{eqnarray}
where $\mu>0$ is a positive constant, and $\xi_i$'s and $\eta_i$'s are i.i.d.\ Gaussian random variables with mean zero and variance one for $i=0,1,\ldots,n$. This corresponds to a one-factor model, where the covariance $\mathbb{E}[(X_i-\mu)(X_j-\mu)]=a^2+J^2c^2$ between assets can partly be explained by the barcode variables ($c\neq 0$). Notice also that all assets are equivalent, i.e.\ the distribution of $(\vec X,\vec Y)$ is invariant under permutations of the assets.

The mutual information on individual assets is given by\footnote{These results can be derived straightforwardly using textbook formulas (see e.g.\ \cite{Cover}).}
\begin{equation}
I(X_i,Y_i) = \frac{1}{2}\log\left(1+J^2\frac{1+c^2}{1+a^2}\right)
\end{equation}
whereas the total information that barcodes provide on $\vec X$ is
\begin{equation}
\label{eq:totinfogauss}
I(\vec X,\vec Y)=\frac{n-1}{2}\log(1+J^2)+\frac{1}{2}\log\left(1+J^2\frac{1+nc^2}{1+n a^2}\right).
\end{equation}
%As we mentioned in the previous section, REMOVED
Since 
\begin{equation}
X=n\mu+\sum_{i=1}^n\xi_i+n a\xi_0+J Y,\qquad Y=\sum_{i=1}^nY_i,
\end{equation}
then the optimal barcode for any $F(X)$ is $G_F(\vec Y)=Y$. Indeed, $I(F(X),\vec Y)=I(F(X),Y)$, which saturates the leftmost inequality in Eq.\ (\ref{dpi:barcodes}). The upper bound on the information content of the barcode is given by
\begin{equation}
I(X,Y)=\frac{1}{2}\log\left(1+J^2\frac{1+nc^2}{1+n a^2}\right).
\end{equation}
We notice that:
\begin{enumerate}
  \item The barcode's information on the portfolio log-return $X$ is larger than that on individual assets (i.e.\ $I(X,Y)>I(X_i,Y_i)$) only if $c>a$, i.e.\ if barcodes are sufficiently informative on the co-movement of assets.
  \item $I(X,Y)$ equals the second term in Eq.\ (\ref{eq:totinfogauss}). Therefore the total loss of information is upper bounded by the first term on the right hand side of Eq.\ (\ref{eq:totinfogauss}), which increases linearly with $n$.
  \item The total loss of information $I(\vec X,\vec Y)-I(X,Y)$ is independent of $a$ and $c$, because all the available information on the co-movement of stocks is captured by $Y$. 
  \item When barcodes are not informative about the correlated variation of assets, i.e.\ for $c=0$, the information content of the barcode $Y$ vanishes $I(X,Y)\to 0$ as $n\to \infty$.
\end{enumerate}

\begin{figure}[htbp!]
\centering
\includegraphics[width=0.48\textwidth]{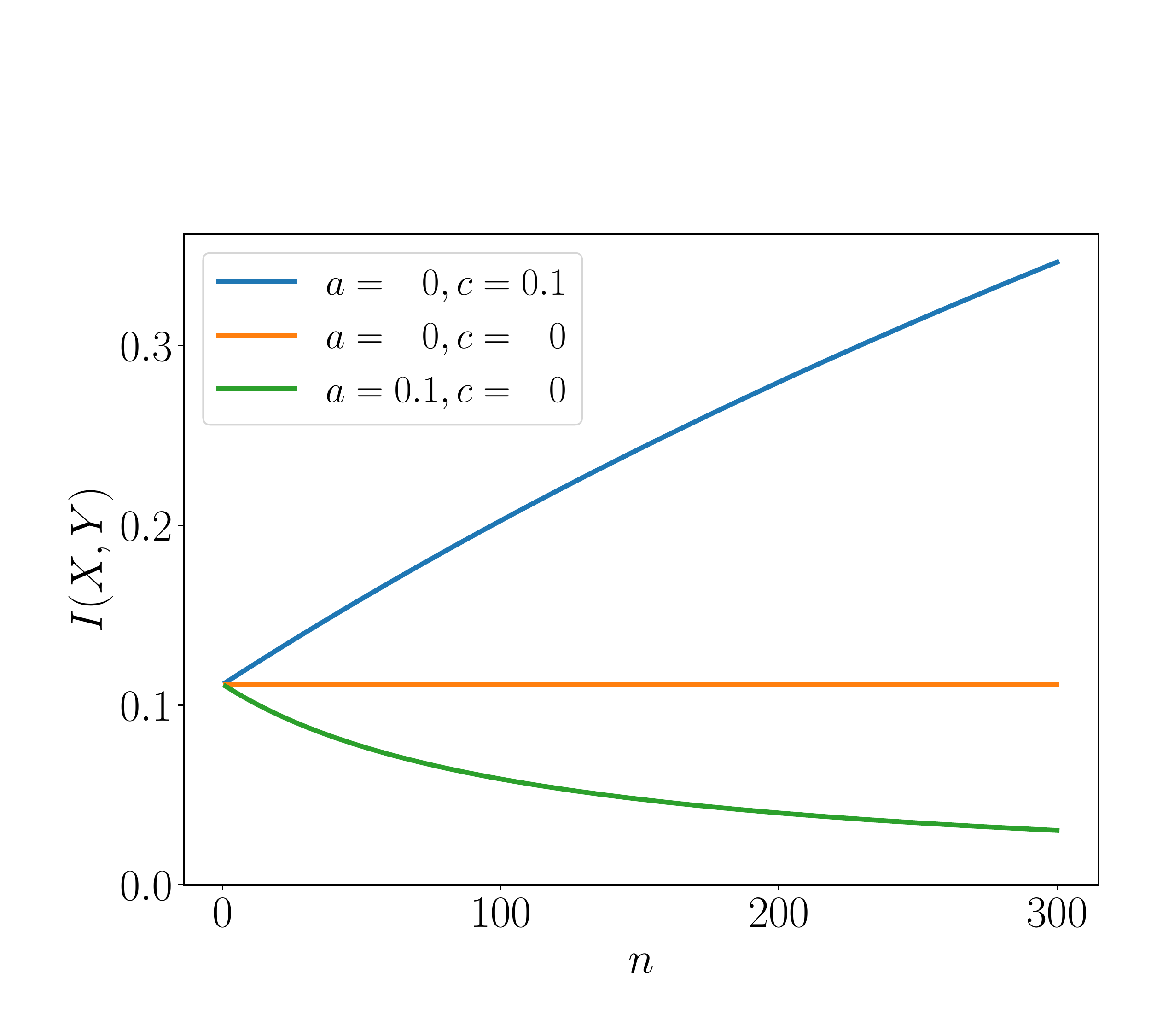}
\includegraphics[width=0.48\textwidth]{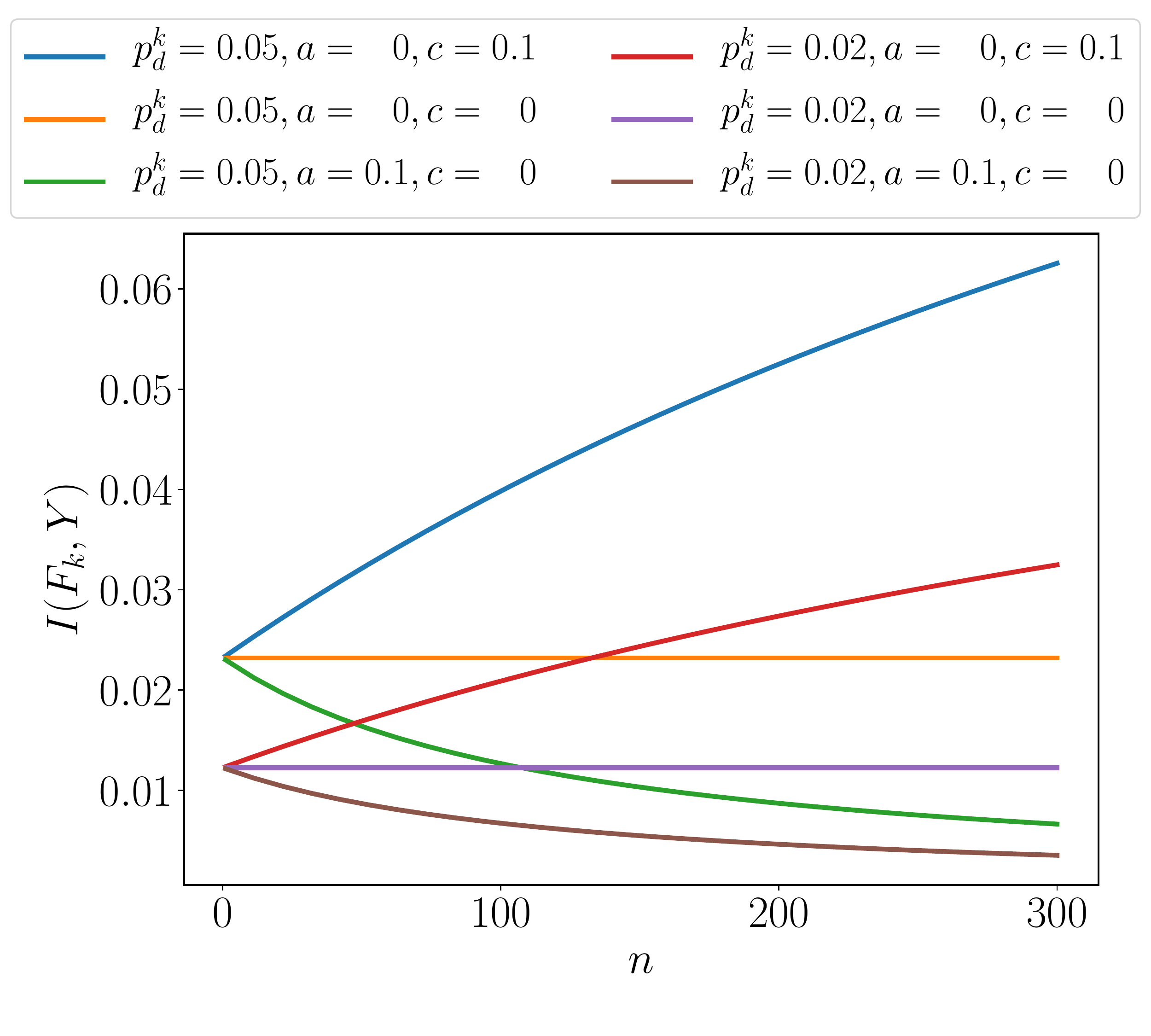}
\caption{Mutual information $I(X,Y)$ and $I(F_k,Y)$ of barcodes for portfolio (left) and ABS (right) generated from Gaussian underlying assets ($J=0.5$), as a function of $n$.}
\label{fig:mi_gauss}
\end{figure}

It is instructive to observe that, when $c\neq 0$, the barcode $Y_j$ provides also information on the return of asset $i$\footnote{A trite calculation shows that
\begin{equation}
I(X_i,Y_j)=-\frac{1}{2}\log\left\{1-\frac{J^2c^2}{(1+c^2)[1+a^2+J^2(1+c^2)]}\right\}.
\end{equation}
}. The case when barcodes are independent ($c=0$) may be appropriate for a portfolio of stocks where $Y_i$ accounts only for fundamental analysis of stock $i$. In this case, very large portfolios become insensitive to information on the fundamentals of individual stocks ($I(X,Y)\to 0$ as $n\to \infty$). This is because $X$ is dominated by the common component $a\xi_0$ on which barcodes $Y_i$ provide no information.
The behaviour of $I(X,Y)$ for the portfolio is summarized in the left panel of Figure \ref{fig:mi_gauss}.

The mutual information can be computed also for the ABS as follows. In a model of gaussian log-returns and information, the threshold parameter $k$ of a tranche $F_k$ can be related to the default probability 
\[
p^k_d= p (X<n\mu+k) = H\left(\frac{k}{\sqrt{\mathbb{V}(X)}}\right)
\]
where $H(z)=\int_{-\infty}^z\frac{dz}{\sqrt{2\pi}}e^{-z^2/2}$ is the cumulative normal distribution function. When information $Y$ is revealed, this default probability changes into
\[
p^k_d(Y)=p (X\le n\mu+k|Y)=H\left(\frac{k-JY}{\sqrt{\mathbb{V}(X|Y)}}\right)
\]
using this expression for $p^k_d(Y)$ and using Eq.\ \eqref{eq:mi},
\begin{equation}
I(F_k,Y)=\mathbb{E}\left[p^k_d(Y)\log\frac{p^k_d(Y)}{p^k_d}+(1-p^k_d(Y))\log\frac{1-p^k_d(Y)}{1-p^k_d}\right],
\end{equation}
where the expectation is taken on the distribution of $Y$.
In the right panel of Figure \ref{fig:mi_gauss} we plot the behaviour of the mutual information for the ABS. $I(F_k,X)$ follows the same qualitative behaviour of $I(X,Y)$ although its value is considerably smaller (more than tenfold in the example of Figure \ref{fig:mi_gauss}). In addition, $I(F_k,Y)$ decreases for safer and safer assets (i.e.\ as $p_d$ decreases), showing that most senior tranches of ABS tend to be remarkably information insensitive.

\subsection{The cost of information: pricing barcodes}

Let us now address the issue of quantifying the value of the information conveyed by the barcodes. The key question we want to address is whether the demand for barcodes can \emph{endogenously} arise in a market. This is possible if barcodes can be priced in such a way that the value of the barcode of a financial instrument provides enough incentives for gathering information on the individual assets. 
We address this question within a mean-variance pricing scheme.
 Hence the setting we consider is that of a portfolio manager that gathers information on $n$ assets, and sells $n$ shares of the resulting portfolio, charging an additional amount related to the price of the information contained in the portfolio's barcode. %\textcolor{blue}{D: The portfolio manager is risk neutral while investors are risk averse\footnote{Banks and large investment funds on the sell side have larger portfolios than small investors on the buy side and can hedge on different risks by taking opposite positions.}.
 	Investors exploit their available information about assets payoffs to price the financial product $Z=F_Z(X)$. When they have access to the barcode, they use the probability $p(Z|Y)$ to asses future performances of the assets, otherwise they use $p(Z)$. In a mean variance framework, the price of $X$ depends on the first two moments of $Z$.
 	When no barcode is provided, the mean variance price reads:
 	\begin{equation}
 	\label{eq:pX}
 	p_Z = \mathbb{E}[Z]-\alpha \mathbb{V}[Z],
 	\end{equation} 	
 	where $\alpha>0$ is the relative risk aversion coefficient. A micro-foundation of the previous formula is discussed in appendix A. As shown there, conditionally on knowing the barcode's value $y$, the price is:
 	\begin{equation}
 	\label{eq:pXy}
 	p_{Z|Y=y} = \mathbb{E}[Z|Y=y]-\alpha \mathbb{V}[Z|Y=y].
 	\end{equation}
 	%Note that the first moment is information insensitive because of the tower property of the conditional expectation: a risk neutral investor ($\alpha=0$) would not value information.\\ 	
Depending on the realized information $y$, the price difference between having or not the barcode can be positive or negative. The price of the barcode should be computed before the realized value of $Y$ is known, therefore it is given by
%Does on average the portfolio manager has an incentive to gather and sell information? Since the portfolio manager is risk neutral she computes the expected price difference. Therefore, the average revenue for selling the barcode is:
 \begin{equation}
 \label{eq:dpz}
 \delta p_Z:=\mathbb{E}[p_{Z|Y}]-p_Z=\alpha\left\{\mathbb{V}(Z)-\mathbb{E}\left[\mathbb{V}(Z|Y)\right]\right\}=\alpha \mathbb{V}\left(\mathbb{E}[Z|Y]\right),
 \end{equation}
 where $\mathbb{V}(Z|Y)$ is the variance of $Z$ on the distribution $p(Z|Y)$ and $\mathbb{E}[Z|Y]$ is the expected value of $Z$ conditional on the value $Y$. Eq.\ (\ref{eq:dpz}) takes the expected value over $Y$ of $\mathbb{V}(Z|Y)$ and the variance of $\mathbb{E}[Z|Y]$ over the distribution of $Y$. 
This result reflects the fact that the knowledge of the distribution of $Y$ does not change the unconditional expected log-return of the asset, but produces a reduction in variance which is equal to the variance of the conditional expected return.

To assess the presence of incentives for barcodes,  we shall compare expected revenues from the barcode with the \emph{cost} of gathering information, which is given by the cost of the barcodes of the original assets. 
%We model the cost of collecting information as (proportional to) the mutual information between the barcode and the associated financial product. 
%We now move on few concrete examples.

When log-returns are given by Eq.\ (\ref{eq:model}), the cost of gathering information for a single asset is 
\begin{equation}
\delta p_{X_i}= \alpha J^2(1+c^2).
\end{equation}
The additional log-return that the optimal barcode ${Y}=\sum_{i}{\eta_i}+nc\eta_0$ yields is 
\begin{equation}
\delta p_{\overline{X}}=\alpha J^2(1/n+c^2).
\end{equation}
Considering this as the price of the barcode that the portfolio manager can charge when selling $\overline{X}$, together with the barcode, we find that the budget's balance for the portfolio manager for selling $n$ shares of $\overline{X}$, is
\begin{equation}
n\delta p_{\overline{X}}-\sum_{i=1}^n\delta p_{X_i}=-\alpha(n-1)J^2
\end{equation}
which is negative. % \footnote{\textcolor{blue}{D: Even allowing the cost of gathering information proportional and not equal to the mutual information, for large $n$ the difference is negative.}}. 
In words, this pricing mechanism does not provide incentives to gather information on individual assets. Interestingly, when barcodes provide information on the co-movement of the assets ($c>0$), the value of information on the whole portfolio, instead, is larger than the sum of the cost of information on individual assets, i.e.\
\begin{equation}
\label{eq:dptot}
\delta p_{X}-\sum_{i=1}^n\delta p_{X_i}=\alpha n(n-1)J^2c^2.
\end{equation}
This is a consequence of the non-linearity of the pricing  mechanism and of the fact that, for $c>0$ the barcode $Y_i$ of asset $i$ provides information also on other asset log-returns $X_j$. Indeed, there is a minimal share size, above which the barcode associated to the log-return $X/m$ provides enough incentives to gather information on individual assets in the sense that $m\delta p_{X/m}-n\delta p_{X_i}\ge 0$. A simple calculation shows that
\begin{equation}
m\le \frac{1+J^2(1+nc^2)}{1+J^2(1+c^2)}.
\end{equation}
The same calculation can be extended to ABS in a straightforward manner, yet the results depends on the way in which the portfolio $X$ is divided into tranches %\textcolor{red}{D: QUESTION: is it possible to have incentives to gather information for some specification of $\{k,F_k\}$? It seems to me that the proof in appendix holds for every choice of $\{k,f_k>0\}$ such that eq. 28 holds true}. If
\begin{equation}
\label{eq:tranch}
X=\sum_k f_k F_k(X)
\end{equation}
%
%\textcolor{blue}{If
%	\begin{equation}
%	\label{eq:tranch}
%	X=\sum_k X f_k F_k(X),
%	\end{equation}
%}
%\textcolor{blue}{where
%	\begin{equation}
%	\label{eq:tranch}
%	F_k= \mathbf{1}_{\{X<k\}}.
%	\end{equation}}
%
%
%\textcolor{blue}{Or, we could define non binary ABS:
%	\begin{equation}
%	\label{eq:tranch}
%	F_k(X):= X \mathbf{1}_{\{X<k\}}.
%	\end{equation}
%}
for some positive constants $f_k$. Then we show in appendix B that
\begin{equation}
\delta p_X> \sum_k \delta p_{f_k F_k(X)}.
\end{equation}
This, together with the fact that for $c=0$ Eq.\ (\ref{eq:dptot}) implies that $\delta p_X=\sum_i\delta p_{X_i}$, shows that if barcodes do not provide information on correlated defaults, securitisation cannot provide incentives to gather information on individual assets (within the present mean variance framework). 
This suggests that, unless information on correlated defaults of individual assets is accounted for, securitisation decreases the value of information contained in barcodes. 

\section{Conclusion}

In this paper we exploit information theoretic concepts to investigate the lack of transparency associated with financial transformations. We discuss a setting where side information about the returns of assets is modelled with an associated random variable, and the information content is quantified using the mutual information. In this setting, we show that every financial transformation implies information losses. In a model of Gaussian log-returns, we find that when fundamental analysis on individual assets is not informative on the co-movement of assets, the information is totally lost in the limit of very large portfolios. In addition, we show that, within a mean variance framework, the value of information also decreases, which suggests that incentives to gather information on individual assets cannot be transmitted across financial transformation. This puts serious doubts of whether market incentives alone are enough to make the introduction of a system of barcodes, as advocated in Ref.\ \cite{Ali2012towards} sustainable. 

These result generalise to a model of assets with binary returns, which is more appropriate for credit derivatives (see Ref.\ \cite{tesiValerio}). The aim of the present paper is that of suggesting ways forward to quantify transparency losses in finance and to raise few key issues. As such, it might be a benchmark for more complex and realistic theoretical models, or for more appropriate schemes to value information in order to overcome these issues.

\bibliographystyle{unsrt}
\bibliography{BibLost}{}

\newpage
\appendix
\section{Mean variance pricing}

In order to do this, we adopt a standard mean-variance framework\footnote{We consider here a setting of incomplete markets, which is appropriate specially for credit markets.}. Consider a representative agent with an initial wealth $W$, 
that is facing the decision of buying $\epsilon W$ units of wealth of an asset with return $Z$. If her utility function is given by $U(\cdot)$, the certainty equivalent $w$ of this investment is defined as that value for which the investor is indifferent between investing in the asset or receiving $w$ units of wealth, i.e.\
\begin{equation}
U(W+w)=\mathbb{E}\left[U(W+\epsilon WZ)\right],
\end{equation}
where $\mathbb{E}[\ldots]$ stands for the expectation on the random variable $Z$.
We take $w$ as a measure of the value of the investment that incorporates the risk premium. Assuming that $\epsilon\ll 1$ and $w\ll W$, we can expand both sides and derive, to leading order, the price of $Z$ as the value per unit of investment.%\footnote{Note that the equation (\ref{eq:pX}) involves only adimensional quantities.}
\begin{equation}
\label{eq:pZ}
p_Z\equiv\frac{w}{\alpha W}\simeq \mathbb{E}[Z]-\alpha \mathbb{V}(Z),\qquad \alpha=-\epsilon\frac{U"(W)W}{2U'(W)}.
\end{equation}
If we further assume investors with constant relative risk aversion (CRRA), then $\alpha$ is a constant, that we assume can be estimated from market data. 
%Consider now the situation where the agent may also access a side information $Y$ at a cost $\delta p_Z$. The value of the investment can again be computed using Eq. (\ref{eq:pX}), and it will depend on $Y$ because expectations should be taken over the conditional distribution $P(Z|Y)$. Yet the value of the information $Y$ needs to be computed {\em before} the value of $Y$ is known, so 
%\begin{equation}
%\label{eq:dpz}
%\delta p_Z=p_{Z|Y}-p_Z=\alpha\left\{\mathbb{V}(Z)-\mathbb{E}\left[\mathbb{V}(Z|Y)\right]\right\}=\alpha \mathbb{V}\left(\mathbb{E}[Z|Y]\right),
%\end{equation}
%where $\mathbb{V}(Z|Y)$ is the variance of $Z$ on the distribution $p(Z|Y)$ and $\mathbb{E}[Z|Y]$ is the expected value of $Z$ conditional on the value $Y$. Eq. (\ref{eq:dpz}) takes the expected value over $Y$ of $\mathbb{V}(Z|Y)$ and the variance of $\mathbb{E}[Z|Y]$ over the distribution of $Y$. 
%This result reflects that fact that the information $Y$ does not change the unconditional expected log-return of the asset, but produces a reduction in variance which is equal to the variance of the conditional expected return.

\section{Pricing ABS}

From Eq.\ (\ref{eq:tranch})
\begin{equation}
\mathbb{E}[X|Y]=\sum_k f_k \mathbb{E}[F_k(X)|Y]
\end{equation}
We assume that $X\ge 0$ and that $f_k>0$. Hence
\begin{eqnarray*}
\delta p_X & = & \alpha\mathbb{V}(\mathbb{E}[X|Y]) \\
 & = & \sum_k f_k^2  \alpha\mathbb{V}(\mathbb{E}[F_k(X)|Y])+\\
 & ~ & \sum_{k\neq k'}\alpha f_k f_{k'}\mathbb{E}\left[
\left(\mathbb{E}[F_k(X)|Y]-\mathbb{E}[F_k(X)]\right)
\left(\mathbb{E}[F_{k'}(X)|Y]-\mathbb{E}[F_{k'}(X)]\right) \right]\\
& > & \sum_k \delta p_{f_k F_k(X)}
\end{eqnarray*}
where the last equation results from the fact that the covariance of $\mathbb{E}[F_k(X)|Y]$ is positive.

\end{document}